\documentstyle[11pt,newpasp,twoside,epsf]{article}
\def\edcomment#1{\iffalse\marginpar{\raggedright\sl#1\/}\else\relax\fi}
\reversemarginpar

\begin{document}
\title{The stellar content of the Galaxy: the white dwarf population}
\author{M. Cignoni, V. Castellani, S. Degl'Innocenti \& S. Petroni }
\affil{Dipartimento di Fisica, P.zza Torricelli 2, 56126, Pisa, Italy}
\author{P. G. Prada Moroni}
\affil{Dipartimento di Fisica, Via Dodecaneso 33, 16146, Genova, Italy}

\begin{abstract}
On the basis of theoretical evolutionary expectations, we develop a
Galactic model reproducing star counts and synthetic color-magnitude
diagrams of field stars, which include the white dwarf (WD)
population. In this way we are able to evaluate the expected
occurrence of WDs in deep observations at the various photometric
bands and Galactic coordinates, discussing the contribution of the WDs
of the various Galactic components. The effects on the
theoretical predictions of different WD evolutionary models, ages,
initial mass functions and relations between progenitor mass and WD
mass are discussed.
\end{abstract}

\section{The model}
Our Galactic model is a three components (spheroid, disc, thick disc)
code which closely follows the ``classic'' Galactic models by Bahcall
\& Soneira (1984, B\&S) and Gilmore \& Reid (1983, see also Paper I
and references therein) concerning the spatial density distribution of
stars.  However, it relies on suitable assumptions on the evolutionary
status and on the initial mass function (IMF) of the various Galactic
populations to reproduce the luminosity functions used as an
(observational) input in Paper I. We recall here the main
characteristics of the model referring, for a detailed discussion, to
Castellani et al. (2001, Paper II).  Predicted results are obtained by
randomly generating star masses according to the adopted IMF (Kroupa
2001) and by using stellar models to derive luminosities in the
selected bands for each given value of the stellar mass and
age. Spheroid stars are assumed to be almost coeval and thus they are
reproduced by populating a suitable theoretical isochrone (age= 12
Gyr, Z= 0.0002, Y= 0.23), while for both the thick disc and the disc
one has to take into account prolonged episodes of star formation.
Thus for these two components star masses and ages are both randomly
generated, the mass distribution reproducing the selected IMF, while a
flat age distribution is adopted within the range assumed for each
component.  For the disc we chose a constant star formation rate (SFR)
from 50 Myr to 9 Gyr to populate evolutionary tracks (Z=0.02 Y=0.27)
until the asymptotic giant branch and the WD cooling sequences. For
the thick disc, following Gilmore, Wyse, \& Jones (1995) and Norris
(1999), we assume a metallicity of $\sim$Z=0.006 and a SFR centered at
$\sim$ 10 Gyr with a spread of few Gyr.  Thanks to this
theoretical approach the model spontaneously predicts the occurrence
of stars in the various evolutionary phases.
\section{The white dwarf population}
By relying on the above described model one can easily predict the
expected abundance of WDs, since each star evolved beyond the AGB
phase and less massive than the lower mass limit for supernovae
(M$_{\rm up}$) is assumed to be a WD. However, to predict the
colour-magnitude (CM) location of these dwarfs one needs further
theoretical ingredients, as given by: i) a WD mass - progenitor mass
relation , ii) theoretical WD models to obtain luminosity and
temperature of a WD as a function of mass and age and iii) suitable
color transformations. As shown in more detail in Paper II, the
prediction on the WD population barely changes if we substitute the
theoretical cooling tracks by Chabrier et al. (2000) with the ones by
Salaris et al. (2000), both transformed into the observational plane
by means of Saumon \& Jacobson (1999) model atmosphere. We checked
also that the predicted WD magnitude distribution shows small
differences changing the semi-empirical WD mass - progenitor mass
relation by Weidemann (2000) with the exponential one by Wood
(1992). In the following we will adopt theoretical cooling tracks by
Salaris et al. (2000) and the semi-empirical mass relation by
Weidemann (2000).  The predicted spheroid WD population obviously
depends on the assumption on the halo age.  Fig. 1 shows the predicted
luminosity functions at four different ages: 6, 8, 10, 13.5 Gyr.  As
well known, for a given age the bulk of the WD distribution is close
but not precisely at the faint end of the cooling isochrone. The
larger is the age the fainter is the bulk of the WD population.  This
is a well understood feature: the increase of the time spent in the
cooling sequence implies a progressive decrease of WD luminosity.
\begin{figure}[hbt]
\plottwo{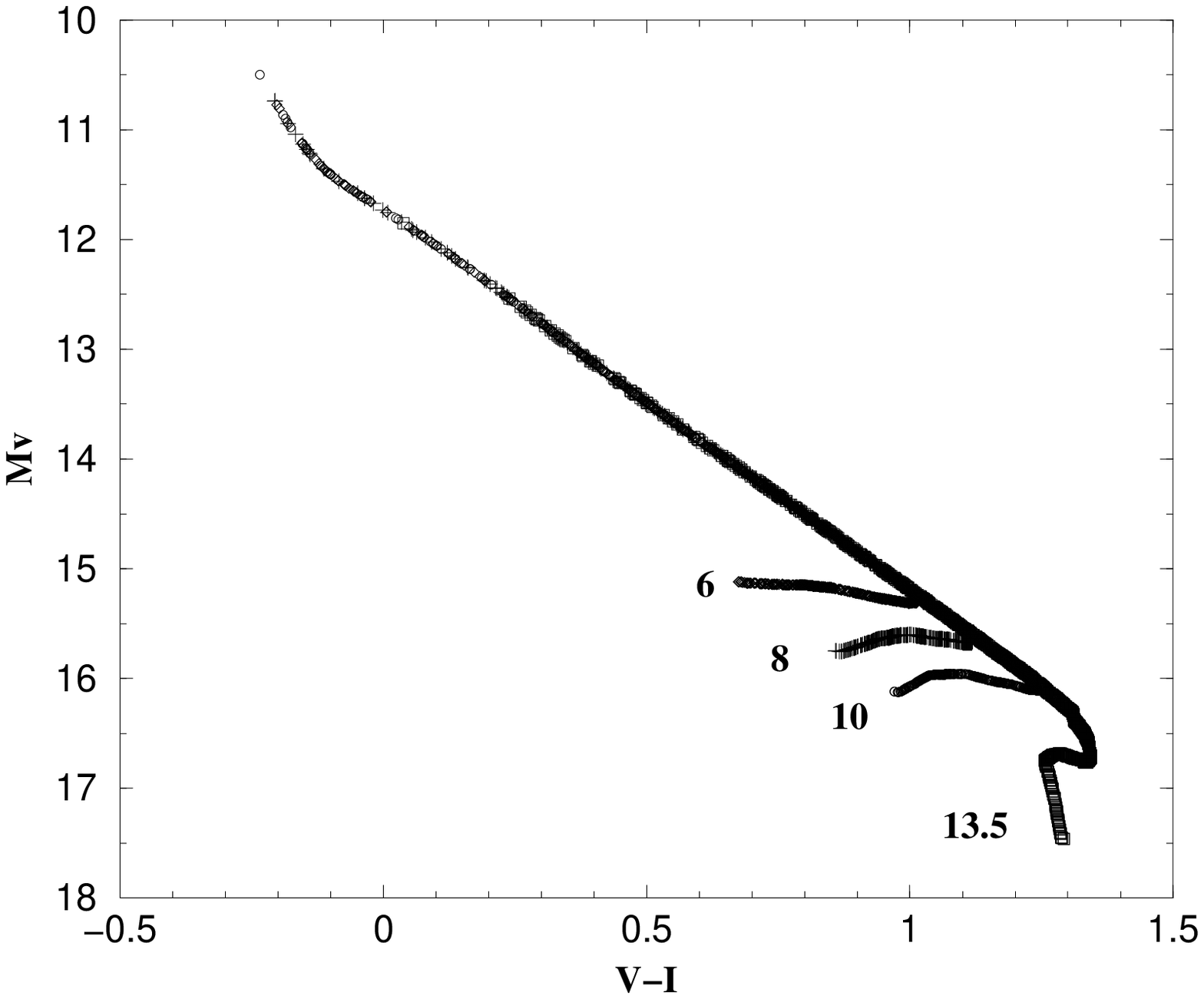}{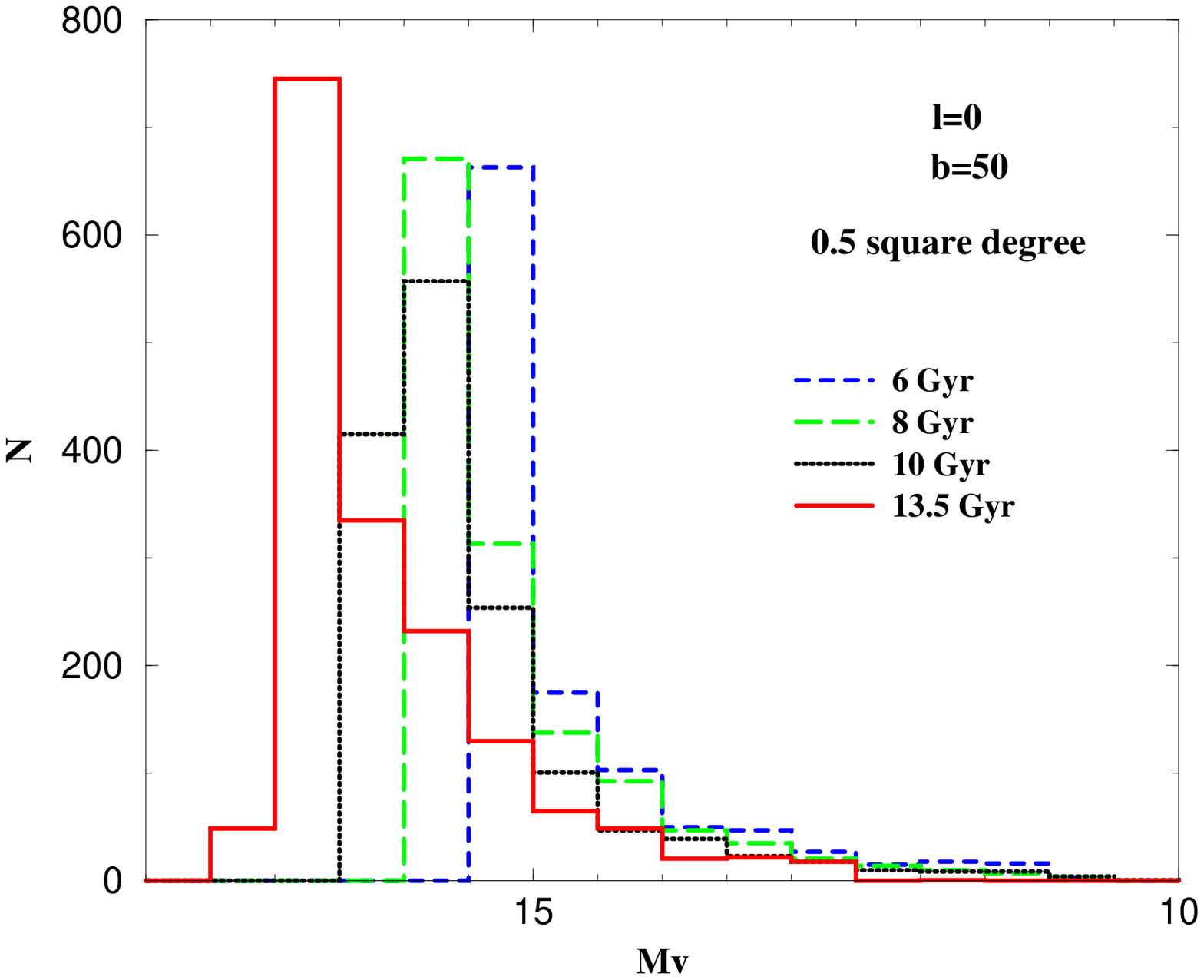}
\caption{Theoretical isochrones in the $M_V-(V-I)$ diagram
for halo WDs of the labeled ages (in Gyr) ({\it left}); predicted
luminosity functions for the spheroid WDs at the labeled ages  ({\it right})
in a field
at l=0\deg, b=50\deg of extension 0.5 square degrees.}
\label{}
\end{figure}
Furthermore the distribution of WD populations significantly depends
on the adopted IMF. The number of WDs is obviously affected only by
IMF variations for masses which could evolve into WDs in a time
shorter than the estimated age of the Universe. 
As range of variation we assume
the uncertainty on the Salpeter exponential ($\alpha$), as evaluated by Kroupa
(2001; $\alpha=2.3\pm0.3$).  As expected, a steeper IMF ($\alpha$=2.6)
depopulates the WD stars. This behaviour can be easily understood as a
consequence of the decrease of the number of stars in the mass range
able to produce WDs.

\section{Results}
Fig 2 and 3 show the predicted results for a field of 0.5 square
degrees at the Galactic coordinates l=0\deg, b=50\deg. 
\begin{figure}[hbt]
\plottwo{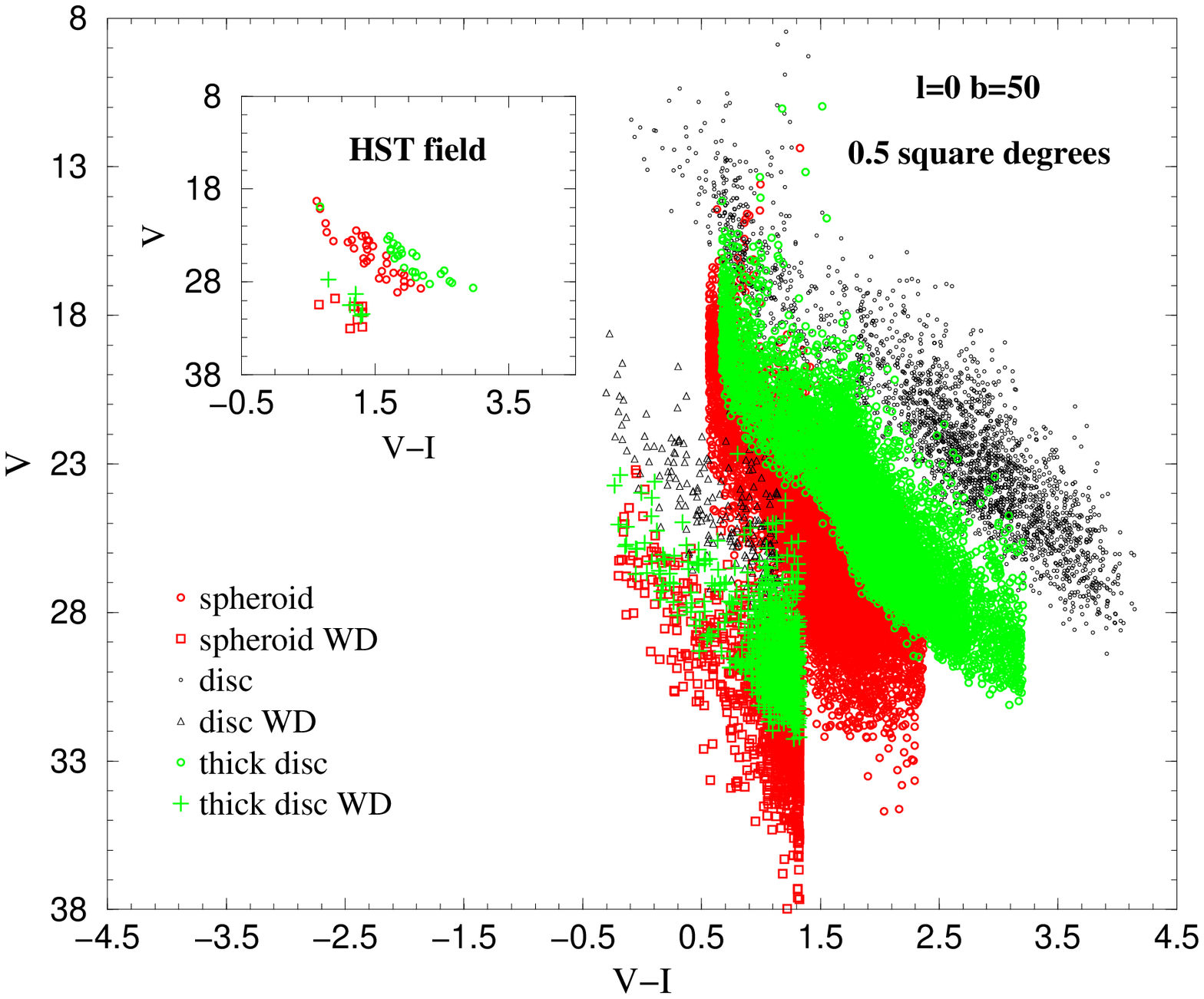}{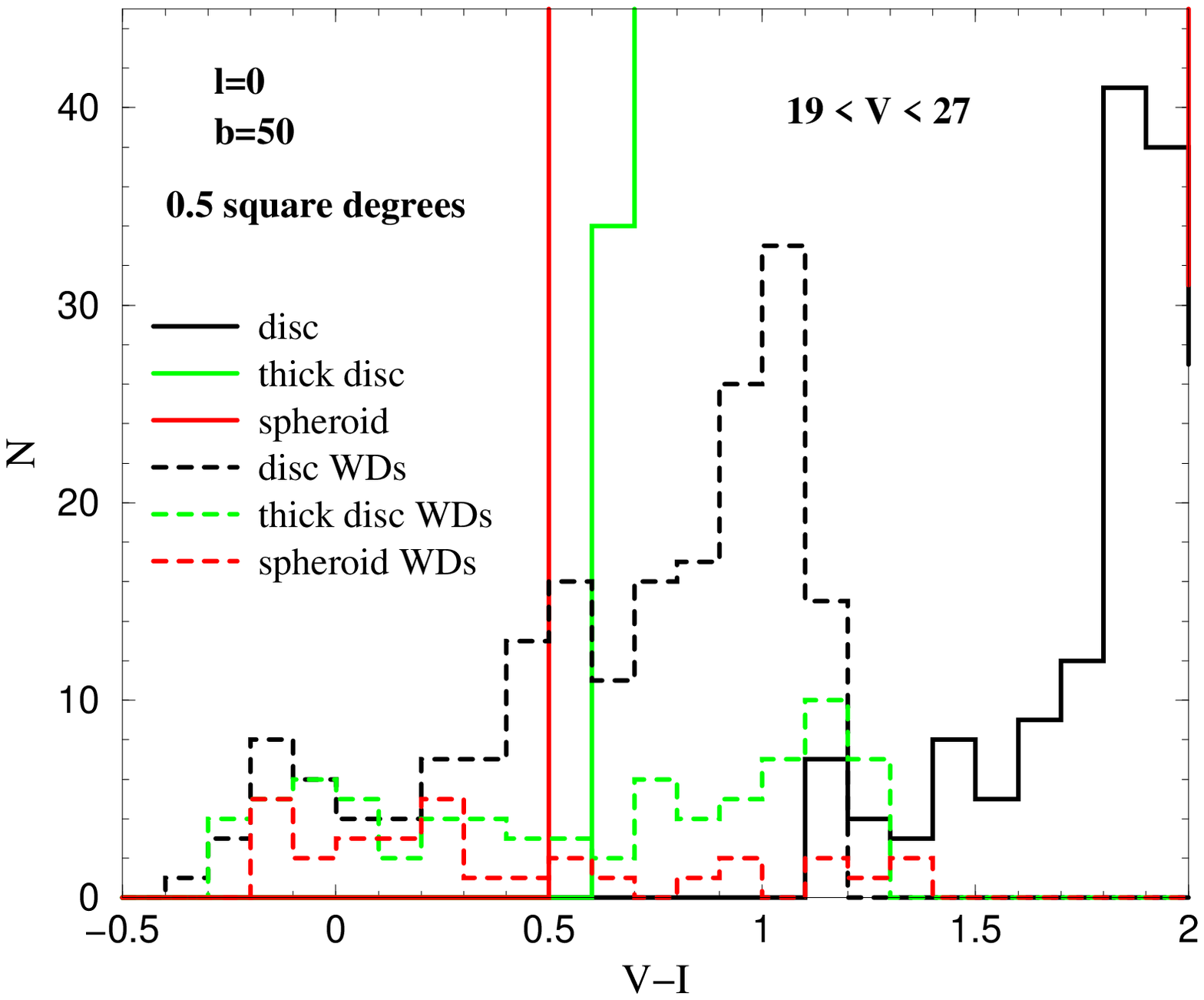}
\caption{Theoretical $(V-I,
V)$ CMD ({\it left}); and ($V-I$)-colour distribution ({\it
right}) for stars in a field at 
l=0\deg, b=50\deg and area of 0.5 square degrees.
 Symbols for the various Galaxy populations as labeled. 
Note that in our model we do not introduce artificial colour
dispersion simulating observative spread in colours.}
\end{figure}
\begin{figure}[hbt]
\plottwo{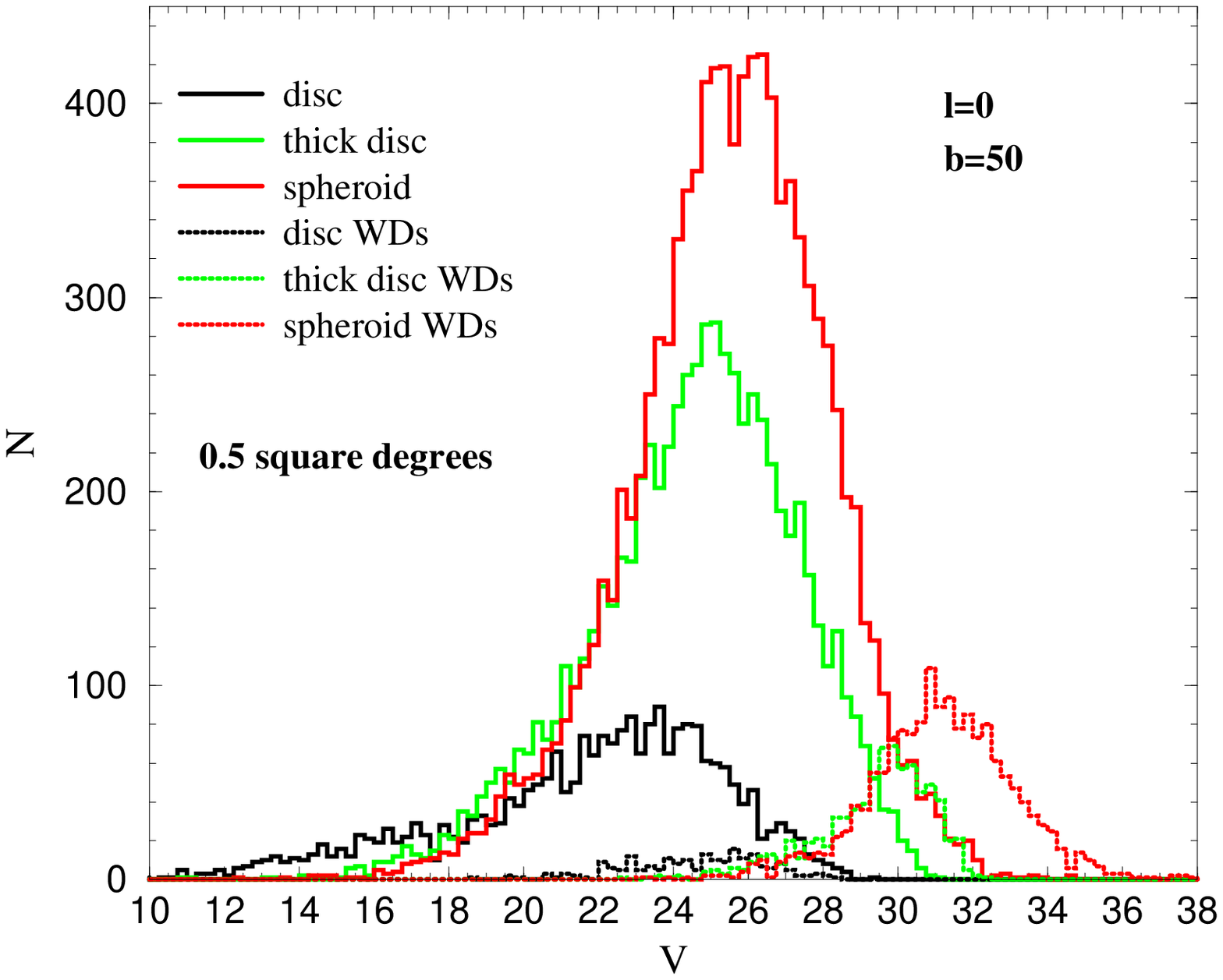}{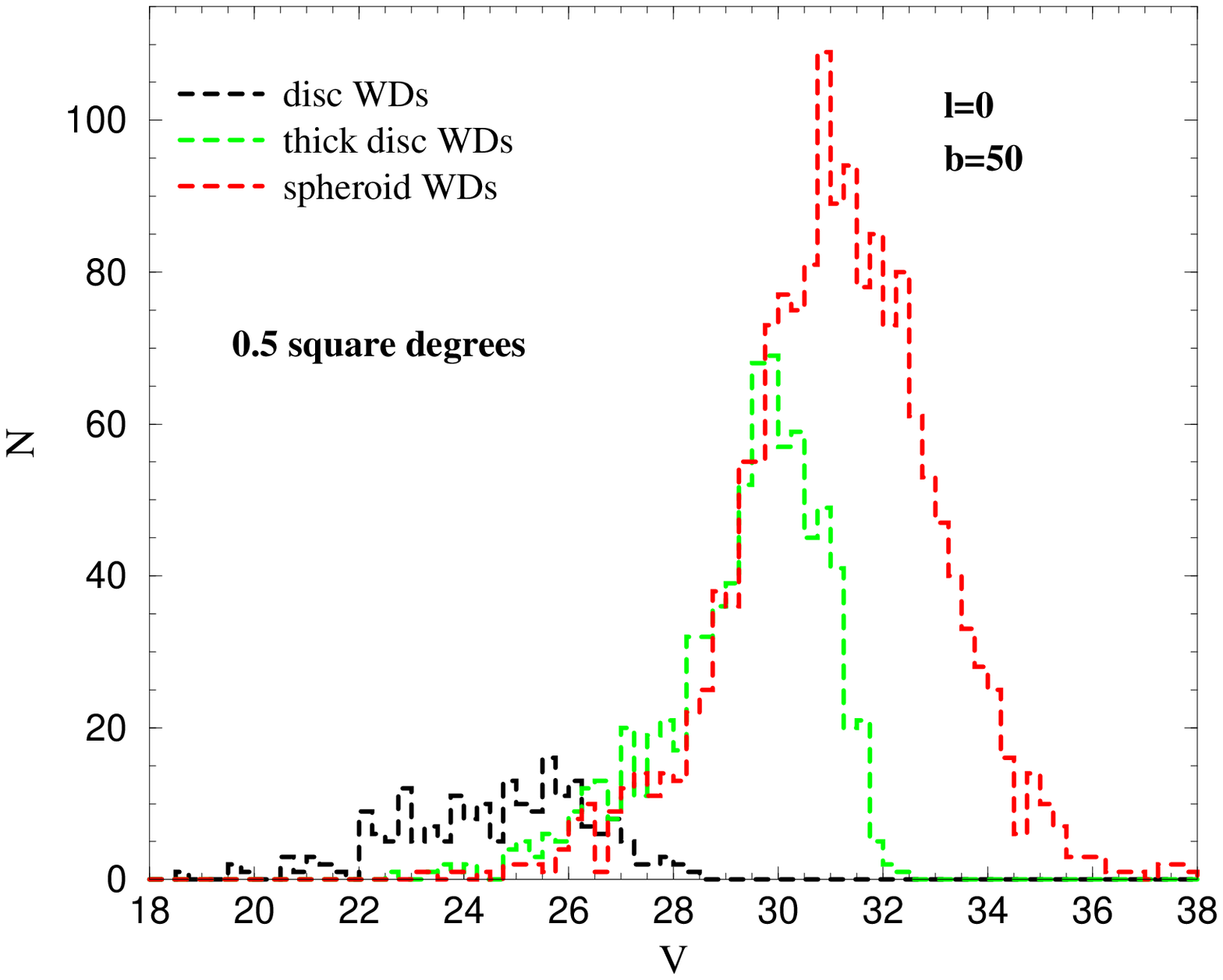}
\caption{Theoretical prediction for field stars in the direction
l=0\deg, b=50\deg and area of 0.5 square degrees.  Theoretical
apparent-$V$ magnitude distribution ({\it left}); zoom of the previous
one in which only WDs are shown ({\it right}).} 
\label{results}
\end{figure}
The small panel in the CM diagram (CMD) (left figure) show the star
counts for a field of about 6.6 arcmin$^2$ of extension, that is the
area generally covered by {\em{Hubble Space Telescope}} ({\em{HST}})
observations (see e.g. King et al. 1998).  Due to the presence of the
thick disc population, the whole sample of stars in the CMD appears
rather uniformly spreaded, without the ``double peaked'' feature
usually predicted by the two-components models because of the
separation in colour between the disc and the spheroid population
(see e.g. B\&S, 1984).  An important finding of our prediction is that
observations down to $V=28$ include almost the whole disc population
and disc WDs.  Only a few percent of thick disc and spheroid WDs are
observable to this luminosities; from Fig. 3 one sees that the thick
disc WD $V$-magnitude distribution appears centered at $V\sim 30$,
while the spheroid distribution is centered at $V\sim 31$, with a tail
reaching faint luminosities down to $V\sim 36$.  From the
($V-I$)-colour distributions (Fig 2, right panel) it appears that the
whole WD population takes place at colours bluer then ($V-I$)$ \sim 1.5$.
Moreover, at sufficiently blue colours, i.e. ($V-I$)$ \la 0.5$, and not
too high luminosity ($V \ga 16$), the sample is constituted
exclusively by WDs.
\section{Conclusions}
We described our galactic model able to reproduce star counts and
synthetic CMDs of field stars from the main sequence to the WD
evolutionary phase at the various photometric bands and Galactic
coordinates. The main goal is the introduction of disc/thick disc and
spheroid WD population in an evolutionary consistent way.  We found
that the WD population is barely sensitive to a change of theoretical
WD models or to a variation of the adopted relation between the WD
mass and the progenitor mass.

\end{document}